\documentclass[12pt]{iopart}

\usepackage{amssymb}
\usepackage{amstext}
\usepackage[cp1250]{inputenc}
\usepackage{iopams}
\usepackage{amsfonts}
\usepackage{eurosym}
\usepackage{epsf}
\begin{document}

\title[A subjective supply-demand]{A model of subjective supply-demand:\\ the
maximum Boltzmann/Shannon entropy solution}

\author{Edward W Piotrowski$^1$ and Jan S\l adkowski$^2$}

\address{$^1$ Institute of Mathematics, The University of Bia\l
ystok, Pl-15424 Bia\l ystok, Poland} \address{$^2$ Institute of
Physics, The University of Silesia, Uniwersytecka 4, Pl-40007
Katowice, Poland} \eads{\mailto{ep@wf.pl},
\mailto{jan.sladkowski@us.edu.pl}}
\begin{abstract}
The present authors have put forward a projective geometry model
of rational trading. The expected (mean) value of the  time that
is necessary to strike a deal and the profit strongly depend on
the adopted strategies. A frequent trader often  prefers maximal
profit intensity to the maximization of profit resulting from a
separate transaction because the gross profit/income is the
adopted/recommended benchmark. To investigate activities that have
different periods of duration we define, following the queuing
theory, the profit intensity as a measure of this economic
category. The profit intensity in a repeated trading has a unique
property of  attaining its maximum at a fixed point regardless of
the shape of demand curves for a wide class of probability
distributions of random reverse transaction (ie closing of the
position). These conclusions remain valid in an analogous model
based on supply analysis. This type of market games is often
considered in the research aiming at finding an algorithm that
maximizes profit of a trader  who negotiates prices with the Rest
of the World (a collective opponent) that posses a definite and
objective supply profile. Such idealization neglects the sometimes
important influence of an individual trader on the demand/supply
profile of the Rest of the World and in extreme cases questions
the very idea of demand/supply profile. Therefore we put forward a
trading model  in which the demand/supply profile of the Rest of
the World induces the (rational) trader to (subjectively) presume
that he/she lacks (almost) all knowledge concerning the market but
his/hers average  frequency of trade. This point of view
introduces maximum entropy principles into the model and broadens
the range of economics phenomena that can be perceived as a sort
of thermodynamical system. As a consequence, the profit intensity
has a fixed point with a astonishing connection with Fibonacci
classical works and looking for the quickest algorithm for
extremum of a convex function: the profit intensity reaches its
maximum when the probability of transaction is given by the Golden
Ratio rule $\frac{\sqrt{5}-1}{2}$. This condition sets a sharp
criterion of validity of the model and  can be tested with real
markets data.

\end{abstract}

\submitto{Journal of Statistical Mechanics -- Theory and
Experiments}
\maketitle

\section{Introduction: profit intensities}
\label{intro}
 The very aim of any conscious and rational economic
activity is optimization of profit in given economic conditions
and, usually, during definite intervals. The interval is usually
chosen so that it contains a certain characteristic economic cycle
(e.g. one year, an insurance period or a contract date). Often, it
is possible but, unfortunately, always risky to make prognoses for
a more or less distant future of an undertaking by extrapolation
from the already known facts, eg by ``sort of statistical
analysis''\footnote{Such a prognosis is often just a disguised
guess, especially if is connected with stock exchange
activities.}. The quantitative description of an undertaking is
extremely difficult when the  duration of the intervals in
question is itself a random variable. The profit gained during a
specific period, described as a function of duration, $\tau$,
becomes also a random variable and as that does not measure the
quality of the undertaking and often its even difficult to find
some reasonable benchmarks for making comparisons. To investigate
activities that have different periods of, possibly random,
duration we define, following queuing theory \cite{queu}, the {\it
profit intensity} as a measure of this economic category
\cite{mmm}. An acceptable definition of the profit must provide us
with an additive function. It seems to us that the  notion of {\it
the interval interest rate} used in this paper leads to consistent
results. Nevertheless, such models, although simple and elegant,
have several drawback from both theoretical and practical points
of view. This type of market games forms often the basis for
research aiming at finding an algorithm that maximizes profit of
an agent who negotiates prices with the Rest of the World (a
collective opponent denoted by RW in this paper) that posses a
definite and objective supply profile. But one cannot claim that
there always is a unique, adversary-independent probability
distribution function, pdf(q), of agents demand or supply profile
expressed as a function of the price $q$. Actually, the RW demand
profile (ie the shape of its pdf(q) curve) in a play against an
agent, say Alice, results from the interaction among all the
agents in question. Alice can probe into the RW demand (or supply)
only by past events analysis\footnote{Note that stock exchange
regulations often allow the possibility of (at least partial)
invisibility of bids. Therefore one cannot be sure of the actual
volume of demand or supply.}. Unfortunately, even a thorough
analysis can produce paradoxical or unwanted recommendations (the
winners curse etc). If such market phenomena are analysed from the
game theory point of view, we have in hands interesting new
(natural?) tools for analysis of paradoxes that follows from
quantum game theory \cite{next}\footnote{"Quantization" often
suggests ways of avoiding paradoxes in game theory due to the
absence of limitations of the classical theory of probability.
This approach has interesting consequences in decision sciences,
cf for example papers by D. Aerts and M. Czachor \cite{czach}, E.
Haven \cite{haven}, A. Yu. Khrennikov \cite{khren},  A. Yu.
Khrennikov and E. Haven \cite{hk}, Piotrowski and S\l adkowski
\cite{newc}, D. Sornette \cite{sor} and others. Of course, we do
not claim that quantum processes play explicit role
 here.}. Such a possibility would be welcome because the non-gaussian shape
of  the demand (supply) curve suggests the existence of Giffen
goods \cite{stig,gif}. Obstacles in "quantization" of such models
can be overcome by replacing the maximum Boltzmann/Shannon entropy
principle with the requirement that the Fisher information gets
its minimum (a discussion on the connection between the principle
of minimum of Fisher information and equations of quantum theory
can be found in \cite{Frieb}). In this way a simple method of
quantum-like  game theory models that stem from statistical
considerations and, what we show bellow, allow us for analysis of
subjectivity in strategy selection. This paper is organized as
follows. First we describe the merchandizing mathematician model
put forward in ref. \cite{mmm} and quote the relevant definitions.
Then we argue for the use of logarithmic quotations and define the
logarithmic rate of return and shortly describe the advantages of
projective geometry approach and scaling invariance of the
resulting models \cite{geom}. The results showing the usefulness
of Boltzmann/Shannon entropy as a measure of strategy quality and
the probability of   making profits will be given in Section 3.
There also the astonishing  emergence of the golden ratio  as a
characterization of the inclinations towards concluding deals of
the most wealthy agents is discussed. Finally we will point some
issues that yet should be addressed.
\section{The  merchandizing mathematician model}
\label{sec:1}

\subsection{The profit intensity}
\label{sec:2} To proceed, let us denote by $t$, $\upsilon _{t}$
and $\upsilon _{t+\tau}$
 the beginning of an interval of the duration $\tau $, the
value of the undertaking (asset) at the beginning  and at the end
of the interval in question, respectively. We  will measure
profits with the help of the {\it logarithmic rate of return}
$r_{t,t+\tau}$ defined as

\begin{equation}\label{rate}
r_{t,t+\tau}\equiv \ln \left( \frac{\upsilon _{t+\tau}}{\upsilon
_{t}}\right).
\end{equation}

The expectation value of the random variable $\xi$ in one trading
cycle (buying-selling or vice versa)  is denoted by $E\left( \xi
\right)$. If $E\left( r_{t,t+\tau} \right)$ and $E\left( \tau
\right)$ are finite,  we can define the {\it profit intensity} for
one cycle $\rho _{t}$  as

\begin{equation}\label{intensity}
\rho _{t} \equiv \frac{E\left( r_{t,t+\tau} \right)}{E\left( \tau
\right)}.
\end{equation}

 This formula is an immediate consequence of the Wald
identity \cite{res}:

\begin{equation}\label{wald}
 E\left( S_{\tau'}\right)= E\left( X_{1}\right)E\left( \tau'\right)\ ,
\end{equation}

where $S_{\tau'}\equiv X_{1}+\ldots +X_{\tau'}$ is the sum of
$\tau'$
 equally
distributed random variables $X_{k},\ k=1,\ldots , \tau'$ and
$\tau'$ is the stopping time \cite{queu,res}. The profit intensity
we have defined in Eq.\ref{intensity} is just the expectation
value of $X_{1}$
 in the Wald identity
Eq.\ref{wald}. The expected profit is the left hand side of the
Wald identity. If we are interested in the profit expected in a
time unit, we have to, according to Wald, divide the expected
profit by the expectation value of the stopping time. This lead to
 the formula given in Eq.\ref{intensity}. We can also calculate
the variance of the profit intensity by using the proposition
10.14.4 from the Resnick's book \cite{res}:

\begin{equation}\label{var}
E\left( \Bigl(S_{\tau'} -\tau' E\left( X_{1}\right)\Bigr)
^{2}\right) =E\left( \tau'\right) Var \left( X_{1}\right) .
\end{equation}
 Note that the above definition of the profit intensity is
applicable also in more general cases when the random variables
$X_{i}$ are correlated or have different distributions.

The expectation value of the profit  during an arbitrary time
interval, say $\left[ 0,T\right]$ is given by the formula

\begin{equation}\label{zysk}
\rho _{0,T} \equiv \int _{0}^{T} \rho _{t}dt \ .
\end{equation}

The proposed definition of the profit intensity is a very
convenient starting point for analysis of various models based on
the subjectivity of demand/supply ides, see the discussed below
models. Relations to the commonly used measures of profits
(returns) can be easily obtained by simple algebraic manipulations
and is omitted here. \subsection{The  merchandizing mathematician
model}\label{sec:3}

The simplest possible market consists in exchanging two goods
which we would call {\it the asset} and {\it the money} and denote
by $\Theta $ and $\$ $, respectively. The  model consist in the
repetition\footnote{In principle, the process is continued
endlessly. } of two simple basic moves:
\begin{enumerate}
\item First move  consist in a rational buying (see below) of the
asset $\Theta $ (exchanging $\$ $ for $\Theta )$.

    \item    The second move
consist in a random  selling of the purchased in the first move
amount of the asset $ \Theta $ (exchanging $\Theta $ for $ \$ $).
\end{enumerate}By rational buying we mean a purchase bound by a
fixed  {\it withdrawal price} $-a$ that is such a logarithmic
quotation for the asset $\Theta $, $-a$, above which the trader
gives the buying up.  A random selling can be analogously
identified with the situation when the withdrawal price is set to
$-\infty $ (the trader in question is always bidding against the
rest of traders). The order of these transactions can be reversed
and, in fact, is conventional. Note that he quotation method does
not matter to the discussed process as long as it can be repeated
many times. Let $V_{\Theta}$ and $V_{\$ }$ denote some given
amounts of the asset and the money, respectively. If at some time
$t$ the assets are exchanged in the proportion $V_{\$}:V_{\Theta}$
then we call the number
\begin{equation}\label{logarytmiczna} p_{t}\equiv\ln \left( V_{\$}
\right) - \ln \left( V_{\Theta} \right)
\end{equation} {\it the
logarithmic quotation} for the asset $\Theta $. If the trader buys
some amount of the asset $\Theta $ at the quotation $p_{t_{1}}$ at
the moment $t_{1}$ and sells it at the quotation $p_{t_{2}}$ at
the later moment $t_{2}$ then his profit (or more precise the
logarithmic rate of return) will be equal to
\begin{equation}\label{logarytmiczna2}
 r_{t_{1},t_{2}}=p_{t_{2}}-p_{t_{1}}.
\end{equation}
The logarithmic rate of return, contrary to $p_{t}$, does not
depend on the choice of unit used to measure the assets in
question. From the projective geometry point of view \cite{geom},
$r_{t_{1},t_{2}}$ is an invariant and $p_{t}$ is not, cf the
discussion of demand and supply curves in Sec.~\ref{sec:4}.
 Let us
suppose now that the model describes a stationary process, that is
the probability distribution  $\eta \left( p \right)$ of the
random variable $p$ (the logarithmic quotation) does not depend on
time. Note that it is sufficient to know the logarithmic
quotations up to arbitrary constant because what matters is the
profit and profit is always a difference of quotations. This is
analogous with the classical physics (eg Newton's gravity) where
only differences of the potential matter. Therefore we  suppose
that the expectation value of the random variable $p$ is equal to
zero, $E\left( p \right) =0$. In addition, we also suppose that
the market is large enough not to be influenced by the activity of
our trader. Let the expression $[sentence]$ takes value 0 or 1 if
the $sentence$ is false or true, respectively (Iverson convention)
\cite{Iverson}. The mean time of a random transaction (buying or
selling) is denoted by $\theta $. The value of $\theta $ is fixed
in our model due to the stationarity assumption. Besides, to
eliminate paradoxes (e.g. infinite profits during finite time
spreads) $\theta $ should be greater than zero. Let $x$ denote the
probability that the rational buying \underline{does not occur}:

\begin{equation}\label{x}
x\equiv E_{\eta}\left( \left[ p>-a\right] \right) .
\end{equation} The
expectation value of the rational buying time of the asset $\Theta
$ is equal to  \begin{equation}\label{oczekiwanie} \theta\left(
\left( 1-x\right) +2x\left( 1-x\right) +3x^{2} \left( 1-x\right)
+4x^{3} \left( 1-x\right) +\ldots \right) .
\end{equation} The ratio of the
expected value of the duration of the whole buying-selling cycle
($\tau$) and the expected time of a random reverse transaction
($\theta$) is given by
$$\begin{array}{rl}\frac{E_{\eta}\left( \tau \right)}{\theta} =& 1 + \left( 1-x\right) \sum
_{k=1}^{\infty} kx^{k-1}\\ =& 1 +  \left( 1-x\right) \frac{d}{dx}
\sum _{k=0}^{\infty} x^{k}\\ =& 1 +  \left( 1-x\right)
\frac{d}{dx} \frac{1}{1-x} = 1+ \frac{1}{1-x}\end{array} .
$$ Therefore the  mean duration of the whole cycle is given
by \begin{equation}\label{cykl} E_{\eta}\left( \tau \right)
=\left( 1 + \left( E_{\eta}\left( \left[ p\leq -a\right] \right)
\right) ^{-1} \right) \theta .
\end{equation} The logarithmic rate of return
for the whole cycle is
\begin{equation}\label{zyskzcyklu}
 r_{t,t+\tau}=- p_{\rightarrow \Theta} +p_{\Theta \rightarrow} ,
\end{equation}
 where the random variable $p_{\rightarrow \Theta}$
(quotation at the moment of purchase) has the distribution
restricted to the interval $(- \infty ,-a]$:
\begin{equation}\label{rozkladp}
\frac{\left[ p\leq -a\right] }{E_{\eta} \left( \left[ p\leq -a
\right] \right) }\; \eta \left( p \right) .
\end{equation}The random
variable $p_{\Theta \rightarrow}$ (quotation at the moment of
selling) has the distribution $\eta$, as the selling is at random.
The expectation value of the of the profit after the whole cycle
is\footnote{$\rho _{\eta}\left( a \right)$ should not be mistaken
for the profit intensity also denoted by $\rho$, cf eq.
\ref{intensity}. They differ by a factor equal to   the (inverse)
average cycle duration.}
\begin{equation}\label{rho} \rho _{\eta}\left( a \right)=
\frac{-\int _{-\infty}^{-a}p\;\eta \left( p\right) dp}{1+ \int
_{-\infty}^{-a}\eta \left( p\right) dp} ,
\end{equation}  which
follows from  Eqs~\ref{zysk} and \ref{zyskzcyklu}. This function
has very interesting properties (we will often drop the subscript
$\eta $ in the following text). First we quote \cite{mmm}:

\begin{flushleft}
{\bf Theorem}\end{flushleft}

The maximal value of the function $\rho ,$ $a_{max},$ lies at a
fixed point of $\rho ,$ that is fulfills the condition

\begin{equation}\label{fixedpoint}
\rho \left( a_{max}\right) = a_{max}.
\end{equation}

 Such a fixed point
$a_{max}$ exists and $a_{max} >0$ \footnote{It tempting to claim
that the function $ \rho $ is a contraction but it is not the
case. Simple inspection reveals that if the probability has a very
narrow and high maximum then $\rho $ is not a contraction in the
vicinity of the maximum. Fortunately, for any realistic
probability density one can start at any value of $a$ and by
iteration wind up in the fixed point (c.f. Banach fixed point
theorem). We skip the details because they are technical and
unimportant for the conclusions of the paper.}.$\Box$

\begin{flushleft}
{\bf Example}\end{flushleft}

For the standard normal distribution with the variance $\sigma$
and expectation value $\hat p$ of a random variable $p$
\begin{equation}\label{przyklad}
\eta \left( p, \sigma \right) \equiv \frac{1}{\sqrt{2\pi}\sigma }
\exp \left( - \frac{\left( p- \hat p \right) ^{2} }{2\sigma
^{2}}\right)
\end{equation}  the expectation value of
the profit during a whole cycle $\rho \left( a, \sigma
\right)_{normal} $ (we have explicitly shown the dependence on the
variance $\sigma $) has a nice scaling property:
\begin{equation}\label{skalowanie}
\rho \left( a, \sigma \right) _{normal}= \sigma \rho \left( a, 1
\right)_{normal} ,
\end{equation}and it is sufficient to work out the
$\sigma=1$ case only. If this is the case we get the maximal
expectation value of the profit for $a=0.27603$. Therefore,
according to Theorem 1, the maximal expected profit is also equal
to $0.27063$. Recall that in our normalization this price is
$e^{0}= 1$. \\

The fixed point theorem recommends the following simple market
strategy  that maximize the trader's expected profit on an
effective market: {\it accept profits equal or greater than the
one you have formerly achieved on average during }
 the characteristic time of transaction
which is, roughly speaking, an average time spread between two
opposite moves of a player (e.g.~buying and selling the same
asset). Unfortunately, such strategy recommendations still
involves some important subjective factors that reflect personal
or even instantaneous attitude towards market state.

\begin{itemize}

 \item  Would a change of information measure for market strategies
result in interesting recommendations?
    \item  Does different measures of
information content results in different trading recommendations?
\item Are there  more useful  (convenient?) information measures
than the Boltzmann/Shannon one?
\end{itemize}
\subsection{The demand and supply curves}\label{sec:4}

The textbooks on economics  abounds in graphs and diagrams
presenting various demand and supply curves\footnote{Blaug
\cite{blaug} quotes at least a hundred of such diagrams.}. This
illustrates the importance the economists attach to them despite
that the whole idea supply/demad profiles has serious drawbacks,
both theoretical and practical \cite{osborn}. Such approach is
also possible within the MM model. To this end let us consider the
functions (the subscripts $s$ and $d$ denote supply and demand,
respectively)
 \begin{equation}\label{Fs}
F_{s} \equiv E_{\eta _{s}} \left( \left[ \xi \leq x\right] \right)
=\int _{-\infty}^{x} \eta _{s}\left( p\right) dp ,
\end{equation} and
\begin{equation}\label{Fd}
F_{d} \equiv E_{\eta _{d}} \left( \left[ \xi \leq x\right] \right)
=\int ^{\infty}_{x} \eta _{d}\left( p\right) dp ,
\end{equation}
where we have introduced two, in general case different,
 probability distribution $\eta _{s}$ and $\eta _{d}$.
They may differ due to the  existence of a monopoly, specific
market regulations, taxes, cultural habits and so on. Let us
recall that  two  methods  of presenting demand/supply curves
prevail in the literature. The first one (the Cournot convention)
is based on the assumption that the demand is a function of
prices. The Anglo-Saxon literature prefers the Marshall convention
with reversed roles of the coordinates. The demand supply or
profile is not not always a monotonic function of prices (c.f. the
discussed below on the turning back of demand/supply curves)
therefore the Marshall convention seems to be less convenient as
one cannot use the notion of a function. The MM model with the
price-like parameter $x$ refers to the Cournot convention.
Therefore, for a given  value $x$ of the logarithm of the price of
an asset $\Theta $, the value of the supply function $F_{s}(x)$ is
given by the probability of the purchase of a unit of $\Theta $ at
the price $e^{x}$. The asset could be provided by anyone who is
willing to sell it at the price $e^{x}$ or lower. The function
$F_{d}(x)$ can be defined analogously. If we neglect the sources
of possible differences between $\eta _{s}$ and $\eta _{d} $ and,
in addition, suppose that at any fixed price there are no
indifferent traders (that is everybody wants to sell or
buy\footnote{That is w we consider only active agents.}) then we
can claim that
\begin{equation}\label{sid} E_{\eta _{s}}\left(  [ \zeta \leq x]
\right) + E_{\eta _{d}}\left( [ \zeta
> x]\right) =1.
\end{equation}

The differentiation of  Eq. \ref{sid} leads to $\eta _{s} =\eta
_{d}$. Under these conditions the price $e^{y}$ for which $E_{\eta
_{s}}\left(  [ \zeta \leq y] \right) = E_{\eta _{d}}\left( [ \zeta
>y ]\right) =\frac{1}{2} $ establishes the equilibrium price (actually,   the most
frequent one).
\subsection{The projective
geometry point of view}\label{sec:5} The model has a natural
setting in the projective geometry formalism \cite{geom}. In this
approach the market is described in the $N-$dimensional real
projective space, $\Re P^{N}$ that is $(N+1)-$dimensional vector
space $\Re ^{N+1}$ (one real coordinate for each asset) subjected
to the equivalence relation $v\sim \lambda v$ for $v\in \Re
^{N+1}$ and $\lambda \neq 0$. That is we assume infinite
divisibility of assets. Actually, a finite field approximation is
possible \cite{K,J} - this problem will be discussed elsewhere.
For example we identify all portfolios having assets in the same
proportions. The actual values can be obtained by rescaling by
$\lambda$. The details could  be found in ref. \cite{geom}. In
this context separate profits gained by buying or selling are not
invariant (coordinate free) but there is an invariant: {\it
anharmonic ratio} of four points. For example for exchange ratio
it gives the relative change of quotation:
$$
{
\{\$,Q,Q',\text{\euro}\}:=\frac{c'_{\$\rightarrow\text{\euro}}}{c_{\$\rightarrow\text{\euro}}}
=\frac{q'_\$ \, q_\text{\euro}}{q'_\text{\euro} \, q_\$}=
\frac{|Q'\text{\euro}|\,|Q\$|}{|Q'\$|\,|Q\text{\euro}|}=}
\frac{P(\triangle_{Q'\text{\euro} O})\, P(\triangle_{Q\$ O})}{
    P(\triangle_{Q'\$O})\, P(\triangle_{Q\text{\euro} O})}
$$The profit $r_{t,t+\tau}$ gained during the whole cycle
is given by the logarithm of an appropriate anharmonic ratio (see
below) is invariant (e.g. its numerical value does not depend on
units chosen to measure the assets). In the space $\Re P^{N}$ this
anharmonic ratio denoted by $[\Theta , U_{\rightarrow \Theta},
U_{\Theta \rightarrow} ,\$]$ concerns the pair of points: $$
U_{\rightarrow \Theta}\equiv\left( \upsilon ,\upsilon \cdot
e^{p_{\rightarrow \Theta }},\ldots \right) \ and \ U_{\Theta
\rightarrow}\equiv\left( w ,w \cdot e^{p_{\Theta
\rightarrow}},\ldots \right)  \eqno(26)$$ and the pair $\Theta \
and \ \$ $. The last pair results from the crossing of the
hypersurfaces $\overline{\Theta}$ and $\overline{\$}$
corresponding to the portfolios consisting of only one asset
$\Theta $ or $\$ $, respectively and the line $U_{\rightarrow
\Theta}U_{\Theta \rightarrow}$. The dots represent other
coordinates (not necessary equal for both points). The line
connecting $U_{\rightarrow \Theta}$ and $ U_{\Theta \rightarrow}$
may be represented by the one-parameter family of vectors
$u(\lambda )$ with $\mu-$coordinates given by
$$u_{\mu}\left( \lambda\right) \equiv \lambda\left( U_{\rightarrow
\Theta} \right) _{\mu} +\left( 1-\lambda \right) \cdot \left(
U_{\Theta \rightarrow} \right) _{\mu} . \eqno(27) $$ This implies
that the values of $\lambda $ corresponding to the points $\Theta
$ and $\$ $ are given by the conditions: $$ u_{0}\left( \lambda
_{\$}  \right) =\lambda _{\$} \left( U _{\rightarrow \Theta}
\right) _{0} + \left( 1-\lambda _{\$} \right) \cdot \left( U _{
\Theta \rightarrow } \right) _{0} =0 \eqno(28) $$ and
$$u_{1}\left( \lambda _{\$}  \right) =\lambda _{ \Theta } \left( U
_{\rightarrow \Theta} \right) _{1} + \left( 1-\lambda _{ \Theta}
\right) \cdot \left( U _{ \Theta \rightarrow } \right) _{1} =0.
\eqno(29) $$ Substitution of Eq. (26) leads to $$ \lambda _{\$}
=\frac{w}{w-\upsilon} \eqno(30) $$ and $$\lambda _{\Theta}=
\frac{we^{p_{\Theta \rightarrow}}}{we^{p_{\Theta \rightarrow}}
-\upsilon e^{p_{\rightarrow \Theta}} }.\eqno(31) $$ The
calculation of the logarithm of the cross ratio $[\Theta ,
U_{\rightarrow \Theta}, U_{\Theta \rightarrow}, \$]$ on the line
$U_{\rightarrow \Theta} U_{\Theta \rightarrow}$ leads to
$$\begin{array}{rl}\ln \left[ U_{\rightarrow \Theta},
U_{\Theta \rightarrow}\right] = &\ln \left[ \frac{we^{p_{\Theta
\rightarrow}}}{we^{p_{\Theta \rightarrow}} -\upsilon
e^{p_{\rightarrow \Theta}} },1,0, \frac{w}{w-\upsilon} \right] \\
= & \ln \frac{v\;w\;e^{p_{\Theta \rightarrow}}}{v\;e^{p_{\Theta
\rightarrow}}\;w } =p_{\Theta \rightarrow} -
p_{\rightarrow\Theta}\end{array} \eqno(32)$$ which corresponds to
the formula (7).
\begin{figure}[h]
\unitlength0.8mm
\begin{center}
\begin{picture}(120,60)
\thicklines \put(80, 0){\line(-4, 3){75}} \put(80,
0){\line(2,5){17}} \put(80,40){\line(-6, 1){75}}
\put(80,40){\line(6,-1){20}} \thinlines \put(80, 0){\line( 0,
1){45}} \put(80, 0){\line( -1, 3){15.5}}
\put(80,40){\line(-2,-5){12.25}} \put(80,40){\line( 4,-3){12.3}}
\put(9,47){$\text{\euro}$} \put(93,40){$\$$} \put(76,42){$Q$}
\put(93,27){$Q_{\$}$} \put(79,-4){$O$}
\put(65,5){$Q_{\text{\euro}}$} \put(60,44.6){$Q'$}
\end{picture}
\caption{Graphical representation of exchange ratios.}
\end{center}
\label{siankoo}
\end{figure}
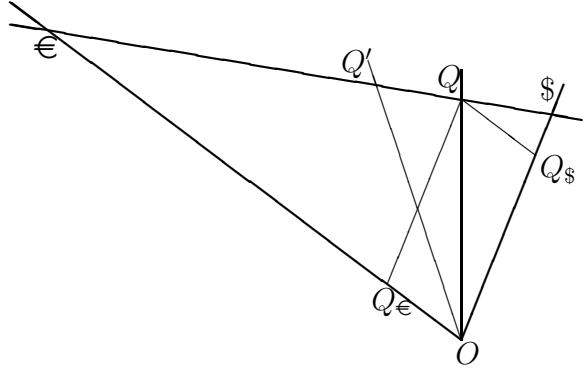
Let us look more closely at  the problem of trading in single
asset. Consider { $$[\mathfrak{G},U_{\to
\mathfrak{G}},U_{\mathfrak{G}\to},\$]
$$} for:
\begin{equation}
\label{wspol-kupiec} U_{\to \mathfrak{G}}:=(v,v\,
\mathrm{e}^{p_{\to \mathfrak{G}}},\ldots)\;\;\text{and }\;\;
U_{\mathfrak{G}\to}:=(w,w\,
\mathrm{e}^{p_{\mathfrak{G}\to}},\ldots)
\end{equation}
 and the points
 $\mathfrak{G}$ i $\$$  given by crossing of the prime
 $U_{\to \mathfrak{G}}U_{\mathfrak{G}\to}$ and one-asset portfolios:
 $\overline{\mathfrak{G}}$ i $\overline{\$}$ corresponding to
assets $\mathfrak{G}$ and $\$$. { The logarithm of the cross ratio
$[\mathfrak{G},U_{\to \mathfrak{G}},U_{\mathfrak{G}\to},\$]$} on
the straight line $U_{\to \mathfrak{G}}U_{\mathfrak{G}\to}$ is
equal to:
\begin{eqnarray}\label{fofo-hkupiec}
 \ln[\mathfrak{G},U_{\to \mathfrak{G}},U_{\mathfrak{G}\to},\$]&=
\ln[\frac{w\, \mathrm{e}^{p_{\mathfrak{G}\to}}}{
         w\, \mathrm{e}^{p_{\mathfrak{G}\to}}-v\, \mathrm{e}^{p_{\to \mathfrak{G}}}},
         1,0,\frac{w}{w-v}]=\\
         &=\ln\frac{v\, w\, \mathrm{e}^{p_{\mathfrak{G}\to}}}{v\,
         \mathrm{e}^{p_{\to \mathfrak{G}}}\, w}= p_{\mathfrak{G}\to }-
         p_{\to \mathfrak{G}}.
\end{eqnarray}
Contrary to the classical economics the balance in the MM model
does not result in  uniform quotations (prices) for the asset
$\Theta $ but only in a stationarity of the supply and demand
functions $E_{\eta _{1}}\left( [\zeta \leq x] \right) $ and
$E_{\eta _{2}}\left( [\zeta \leq x] \right) $. Therefore the MM
model is not valid when the changes in the probabilities happens
during periods shorter or of the order of the mean time
transaction $\theta$. Fortunately,  the presented above stochastic
interpretation of the supply and demand profiles of agents remains
valid in such situations. Moreover, we can consider piecewise
decreasing functions $F_{s}$ and  $F_{d}$. These function cease to
be probability distribution functions because their derivatives
(probability densities) are not positive definite. This
generalization corresponds to  the effect of {\sl turning back} of
the supply and demand curves what often happens for work supplies
and, in general, for the so called  the Giffen goods
\cite{stig,gif}. In the Marshall convention these curves are not
diagrams of functions at all and in the Cournot convention these
curves are diagrams of multivalued functions. Note that in this
way not positive definite probability densities (eg. Wigner
functions) gain interesting economics reason for their very
existence\cite{Feynman}.

\subsection{Strategy selection }
\label{sec:9} By a choice of stochastic process consistent with
the MM model one can determine the dynamics of such a model, c.f.
Ref. \cite{Blaquiere}. Therefore we suspect that the departure
from the laws of supply and demand might be the first known
example of a macroscopic reality governed by quantum-like rules
\cite{qf}. Such hypothetical quantum-like economics could have
started with the evidence given by Robert Giffen in the British
Parliament \cite{stig} and actually could have earlier origin than
the quantum physics. It should be noted here that from the quantum
game theory point of view the Gauss distribution function is the
only supply (demand) curve that fulfills the physical
correspondence principle. The authors would devote a separate
paper to this very interesting problem.

Let us note that the distribution functions allow for correct
description of the famous Zeno paradoxes (when grains form a pile?
when you start to be bald?) because the introduction of
probabilities removes the original discontinuity. For example the
problem of morally rightness of prices: if the price is low (state
0) nobody wants to sell and if the price is high (state 1)
everybody wants to sell. Without the probability theory we are not
able to describe intermediate states which, in fact, are typical
on the markets. Does it suggest that the MM model can also be
applied to problems where there is a necessity of finding maximum
(minimum) of a profit intensity like parameter?

\section{Subjective character of market demand}

The probability distribution $\eta(p)$ that describes the Rest of
the World strategy in the game against a single agent (Alice) is
 de facto of transcendental nature: agents finalizing a deal can
only observe (measure) its results as  values of execution prices
in a way that harks back to Plato's distinction between the
idealized form of a thing and its imperfect realization in our
world as if by a shadow of the form in Plato's Cave. They might
erroneously think that the shadows are ”real” and not just
projections of the outside world. Note that the problem is common
\cite{polyakov}. Therefore, as statistical physics and Shannon's
information theory teach us, the best they can do is to
approximate $\eta(p)$ be the appropriate Gibbs distribution
following from the minimum of their knowledge (information), that
is the maximum of their information entropy, constrained by the
known market parameters. In the analysis of the effectiveness of
strategies the distribution function $\eta(p)$ has to be replaced
by its "shadow". Let us denote the  "shadow" distribution function
by $pdf(p)$. The key problem is that the concrete form of $pdf(p)$
strongly depends on the adopted model of information gathering,
data mining and information measures being used. In this paper we
propose to select Boltzmann/Shannon information entropy $S$ to
this end. Of course, there are other interesting alternatives and
the conclusion concerning the shape of   $\eta(p)$) might differ
in a dramatic way! We will refer to the class of assumptions
regarding
 market, measures of information, data mining,  utility functions and so on adopted by the agent (Alice) in the process of determining $pdf(p)$
 as Alice's imaginoscope (Alice's wall in Plato's Cave).

\subsection{The model}
\begin{figure}[h]
\begin{center}
\epsfysize=60mm
\epsfbox{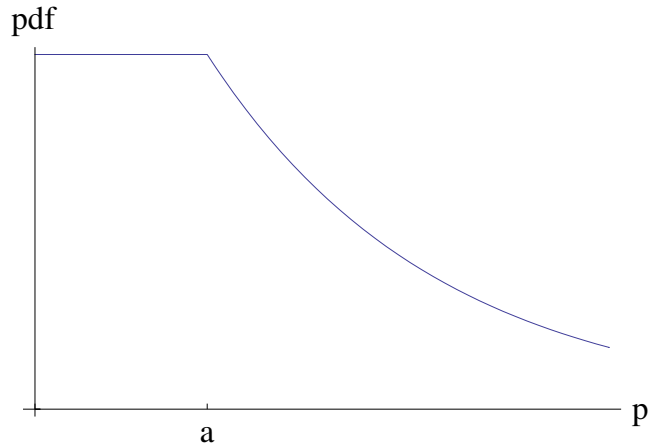}
\end{center}
\caption{\label{label11}The subjective probability distribution
function of market prices.}
\end{figure}

\begin{figure}[h]
\begin{center}
\epsfysize=60mm
\epsfbox{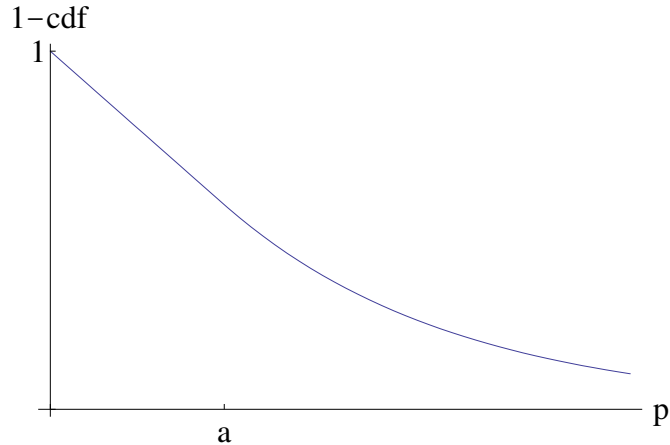}
\end{center}
\caption{\label{label12} The market demand function as
(subjectively) perceived by Alice (the supplier).}
\end{figure}

Consider the simplest case when the domain of $pdf(p)$ one-side
bounded. The assumption that the selling prices are bounded below
(and buying prices bonded above) could by justified by, for
example, the principle of rational production (one does not sell
at prices below production costs), the existence of minimal
salary, regulation concerning usury etc. In addition, let us
assume that Alice uses the minimal number of estimates (that is
one) for $pdf(p)$. The obvious one is the probability of making
the deal $P:=E([p-a])=\int_a^\infty pdf(p) dp$  (Iverson
convention). As we assumed that the support of $pdf(p)$ is bounded
below, say by 0, for an effective reconstruction of the measure
$pdf(p)dp$ she needs at least one unbounded observable. Therefore,
instead of the used in the quantum-like approaches variables
$[q-\psi]$ \cite{qbg,qmg} we will use the observable of surpassing
the minimal return rate $a$ for Alice $(p-a)[p-a]$. Under these
assumptions we get the following Gibbs-like distribution function
for the shadow random variable $p\in [0,\infty)$:
\begin{equation}
pdf(p)=\frac{e^{-\frac{[p-a](p-a)}{T}}}{a+T}\,. \label{gibs}
\end{equation}
For this specific distribution, it follows that the Lagrange
multiplier $T^{-1}$ with the value fixed for the expectation value
$P$ takes the functional form of the logarithm of the withdrawal
price $a$
  scaled by the factor
  equal to the ratio of average number of transactions and the
  average number of transaction that have not been finalized:

$$
T^{-1}= \frac{P}{1-P}\, a\,.
$$
Plot of the function Eq.~$\ref{gibs}$ is presented in
Fig.\ref{label11}.
According to the Laplace indifference principle,  for prices up to
the value $a$ the probability of closing the deal is constant
(Alice does not sell) and for higher the probability distribution
is exponential. The corresponding market demand function
(according to Alice's imaginoscope!) is presented in
Fig.\ref{label12}.
Note that we can observe (see Fig.\ref{label12}) that the
localization of $a$ becomes obscure due to the "continuous
character" of the line tangent to the plot. The demand decreses
 monotonously as prices approach $a$ and then exponentially!a strategy that maximizes
profit. By inserting $pdf(p)$ into Eq. [\ref{rho}] as the
probability distribution $\eta$ and then into the formula
[\ref{fixedpoint}]
 for  the
fixed point of $\rho$ we obtain the condition  under which Alice
profit is maximal:

\begin{equation}
\frac{\int_a^\infty p\, pdf(p)dp}{1+ \int_a^\infty pdf(p)dp}=a\,.
\label{suro}
\end{equation}
After integration and simple manipulations the condition
$(\ref{suro})$ reduces to the very simple formula for the
probability of making transactions:
$$
P=\varphi:=\frac{\sqrt{5}-1}{2}\approx 0.618\,.
$$
In the case of a bounded support of the demand the probability $P$
of making an optimal  transaction for the most profitable (golden)
Alice strategy is about 0.62. Due to the obvious symmetry of the
model with respect to the involution $\Theta\leftrightarrow\$ $,
the solutions and conclusions for the Rest of the World supply and
Alice demand profiles are analogous.

\subsection{Measures of information content of marker shadows}
\begin{figure}[h]
\begin{center}
\epsfysize=60mm
\epsfbox{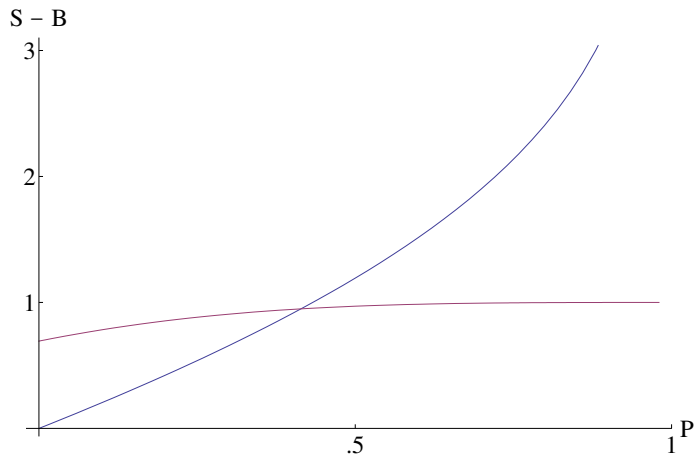}
\end{center}
\caption{\label{label13}Plots of functions  $P-\ln(1-P)$ (blue) and $P-\ln\frac{1+P^2}{2}$ (red).}
\end{figure}

\begin{figure}[h]
\begin{center}
\epsfysize=60mm
\epsfbox{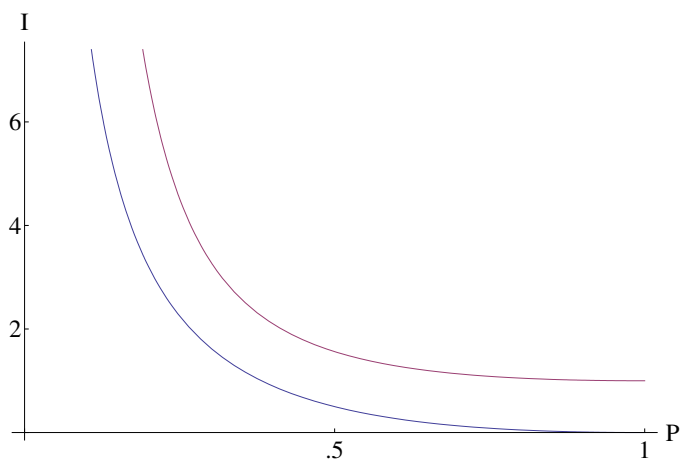}
\end{center}
\caption{\label{label14}Plots of functions $\frac{(1-P)^2}{P}$ (blue) and $\frac{(1+P^2)^2}{(2p)^2}$ (red).}
\end{figure}

\begin{figure}[h]
\begin{center}
\epsfysize=60mm
\epsfbox{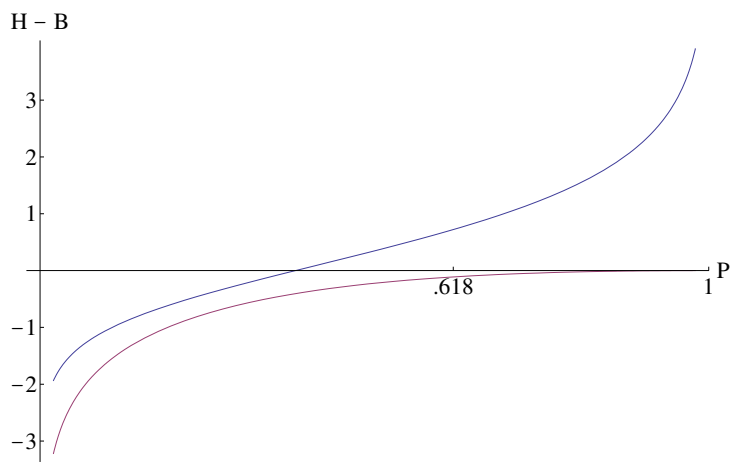}
\end{center}
\caption{\label{label15}Plots of functions  $\ln\frac{\sqrt{P}}{1-P}$ (blue) and $\ln\frac{2 P}{1+P^2}$ (red).}
\end{figure}

If we apply the (continuous form) formula for  Boltzmann/Shanonn entropy
$$S=-\int_0^\infty \ln(pdf(p))\,pdf(p)\,  dp$$ to  subjective projection of market supply probability distribution $(\ref{gibs})$
we obtain expression for the entropy of the knowledge about market
$S(P)$ gathered by the agent, say Alice, during transactions
initiated by  her strategy
  $a$. Fig.~\ref{label13} presents plots of the function $S-B$, the relative entropy $S$
  corrected by the terms $B=-\ln a$ (blue) and $B=-\ln E(p)$ (red), respectively. The market entropy calculated modulo the expectation value
  of the logarithmic return $\ln E(p)$ varies in an unimportant way in the whole domain of $P$.
  This is not accidental. We have already advocated \cite{mmm,geom} advantages of using logarithms of mean rates.
  We see that it a quite good approximation as a measure of information available about  markets.
Unfortunately, the relative entropy B/S $S-B$ has one unpleasant feature: depending on the choice of the parameter (constraint),
substantial changes could be observed c.f. Fig.~\ref{label13}. Nevertheless, abiding by this measure of information,
we can use Fisher information for information content valuation while selecting strategies (that is determining the subjective form of $pdf(p)$):
\begin{equation}
I:= E((\frac{\partial \ln pdf(p)}{\partial p})^2)=\int_0^\infty (\frac{\partial \ln pdf(p)}{\partial p})^2\, pdf(p)\, dp \,.
\label{fisza}
\end{equation}
Plots of $I(P)$ modified by the corresponding factors (i.e. units) $a^{-2}$ (blue line) and $(E(p))^{-2}$ (red line) are presented in Fig.~\ref{label14}.
Now, both curves became similar. The observed decrease in $I(P)$ while $P$ increases is caused by dominating contribution of the integral
over the ignored by Alice domain $[0,a]$ to expression $(\ref{fisza})$, where the probability distribution $pdf(p)$ is constant. Acceptation of
Fisher information (\ref{fisza}) as the measure of information results in smaller susceptibility to subjective changes in attitudes.
This situation suggests that a completely new approach based on Fisher information might be more appropriate. Such approach is described in Ref.~\cite{FBS}. Note that it is possible to modify formula $(\ref{fisza})$ for information $I$  (general $N$-dimensional case):
\begin{equation}
\label{zub}
H:= -\frac{1}{2}\sum_{k=1}^N \ln (E((\frac{\partial \ln pdf(p_1,...,p_N)}{\partial p_k})^2))
\end{equation} to get a new entropy measure that is consistent with Fisher information $I$ in the sense that it generates the
same extrema of $pdf(p)$. But, in addition, it has the property of
being sum of two terms of which the first depends only on $a$ or
$E(p)$ (constraint entropy) and the second depends only on the
probability $P$. This constraint entropy is identical to the
Boltzmann/Shanonn constraint entropy. We present in
Fig.~\ref{label15} the curves that correspond to plots given in
Fig.~\ref{label13}. The model based on maximization of
Boltzmann/Shannon entropy seems to be reliable below the extremal
value $P=\frac{\sqrt{5}-1}{2}\approx 0.618$ (c.f.
Fig.~\ref{label15}) but for market games with higher probabilities
of buying by WR we envisage that the model based on extremum of
the entropy $(\ref{zub})$ would be better.

\section{Conclusions}
Searching optimal solutions and fixed points are the key issues of
contemporary mathematical economics and finance \cite{Deb}. Such
classical results as generalized Brouwer theorem \cite{kak} and
the Brown-Robinson iteration procedure \cite{rob} are widely
applied and useful tools. The model presented in this paper model
combines both ideas with the information theoretical approach. The
 extension of the MM model to the randomized
withdrawal price cases which might also generalize the results of
\cite{acta,13} where thermodynamics of investors was considered
and the temperature of portfolios was defined.  The emergence of
the golden ratio  as a  characterization of the inclinations
towards concluding deals of the most wealthy agents is
astonishing! This condition sets a sharp criterion of validity of
the model and  can be tested with real markets data. Is this
simply a coincidence or there is a deeper connection between these
algorithms? There is no obvious answer to this question. The
golden ratio emerges also in the most efficient algorithm for
finding extremum of convex functions on a segment
\cite{Kiefer,voro}. It turns out that in the market game, the
biggest of games, we can find traces of the golden ratio, so
abundant in other phenomena: an excellent source of bibliography
can be found in The Fibonacci Quarterly, the official journal of
the Fibonacci Association.

\vspace{5mm} {\bf Acknowledgments} This work was supported  by the
scientific network {\it Laboratorium Fizycznych Podstaw
Przetwarzania Informacji}\/ sponsored by the {\bf Polish Ministry
of Science and Higher Education} (decision no
106/E-345/BWSN-0166/2008).

\end{document}